\documentstyle[epsfig]{aipproc}

\begin{document}
\title{Single Spin Asymmetries in Semi-Inclusive  
Electroproduction: Access to Transversity}

\author{\underline{K.A.~Oganessyan}$^{* {\dagger}}$\thanks{E-mail: kogan@hermes.desy.de}, N.~Bianchi$^*$, 
E. De~Sanctis$^*$, W.-D.~Nowak$^\ddagger$ } 
\address{$^*$INFN-Laboratori Nazionali di Frascati, 
via Enrico Fermi 40, I-00044 Frascati, Italy \\
$^{\dagger}$DESY, Notkestrasse 85, 22603 Hamburg \\ 
$^{\ddagger}$DESY Zeuthen, Platanenallee 6, D-15738 Zeuthen, Germany 
}

%\lefthead{LEFT head}
%\righthead{RIGHT head}

\maketitle

\begin{abstract}
We discuss the quark transversity distribution function and a possible way to access 
it through the measurement of single spin azimuthal asymmetry in semi-inclusive 
single pion electroproduction on a transversely polarized target. 

\end{abstract}

At leading order in $1/Q$, the cross section for a hard scattering process is 
given by the convolution of a hard part and a soft part. The former describes the scattering 
among elementary constituents and can be calculated perturbatively in the framework 
of QCD. The latter accounts for the processes in which either partons are produced 
from the initial hadrons (parton distribution functions) or final hadrons are produced from 
partons (parton fragmentation functions) which result from the hard elementary scattering. 

For every quark flavor, besides the well-known parton distribution $f_1(x)$ and the 
longitudinal spin distribution $g_1(x)$, there is a third twist-two distribution 
function, the transversity distribution function $h_1(x)$ which was first discussed by Ralston and 
Soper\cite{RS} in double transverse polarized Drell-Yan scattering. The transversity distribution 
$h_1(x)$ measures the probability to find a transversely polarized quark in a transversely 
polarized nucleon. It is equally important for the description of the spin structure of nucleons 
as the more familiar function $g_1(x)$; their information being complementary. In 
the non-relativistic limit, where boosts and rotations commute, $h_1(x) = g_1(x)$; then  
difference between these two functions may turn out to be a measure for the relativistic 
effects within nucleons. On the other hand, there is no gluon analog on $h_1(x)$. 
This may have interesting consequences for ratios of transverse to longitudinal 
asymmetries in polarized hard scattering processes (see e.g. Ref.~\cite{JJJ}).   

The transversity distribution $h_1(x)$ remains still unmeasured. The reason is that it is a 
chiral odd function, and consequently it is suppressed in inclusive deep 
inelastic scattering (DIS)~\cite{JJ}. Since electroweak and strong interactions 
conserve chirality, $h_1(x)$ cannot occur alone, but has to be accompanied by a 
second chiral odd quantity. 

In principle, transversity distributions can be extracted from cross 
section asymmetries in polarized processes involving a transversely 
polarized nucleon. In the case of hadron-hadron scattering these asymmetries 
can be expressed through a flavor sum involving a product of two chiral-odd transversity 
distributions. This is one of the main 
goals of the spin program at RHIC \cite{RHIC}. An evaluation of the corresponding
asymmetry was carried out \cite{OM} by assuming the saturation of 
Soffer's inequality \cite{SOF} for the transversity distribution: the maximum 
possible asymmetry at RHIC energies was estimated to be about $2 \%$. 
At smaller energies ($\sqrt{s} \simeq 40$ GeV), e.g. for a possible fixed-target 
hadron-hadron spin experiment at the proposed HERA-$\vec N$ facility~\cite{HERAN} 
the asymmetry is expected to be higher (about $4 \%$). 

In the case of {\it semi-inclusive} deep inelastic lepton scattering\footnote{The relevant 
kinematics is: $Q^2=-q^2$ where $q=k_1-k_2$, $k_1$ ($k_2$) being the 4-momentum of the incoming (outgoing) 
charged lepton is the 4-momentum of the virtual photon. $P$ ($P_h$) is the momentum of the 
target (final hadron), $x=q^2/2(Pq)$, $y=(Pq)/(Pk_1)$, $z=(PP_h)/(Pq)$} (SIDIS) 
off transversely polarized nucleons there exist several methods to access 
transversity distributions. One of them, the twist-3 pion production 
\cite{jaffe-ji93}, uses longitudinally polarized leptons and measures a double spin 
asymmetry. The other methods do not require a polarized beam, and rely on the  
{\it polarimetry} of the scattered transversely polarized quark. They consist on: 
\begin{itemize}
 \item the measurement of the transverse polarization of $\Lambda$'s in the 
       current fragmentation region \cite{ArtruMek,jaffe96},
 \item the observation of a correlation between the transverse spin vector
       of the target nucleon and the normal to the two-meson plane 
       \cite{jaffe97b,jaffe97a},
 \item the observation of the ``Collins effect'' in quark fragmentation through 
       the measurement of pion single target-spin asymmetries
       \cite{COL,AK,TM}.  
\end{itemize}          
In the following we will mainly focus on the last method. 

To access the transversity in SIDIS  
off transversely polarized nucleons, one can measure the azimuthal angular dependences 
in the production of spin-0 or (on average) unpolarized hadrons. This production is described 
by the intrinsic transverse momentum dependent fragmentation function $H_1^{\perp}(z)$ 
which is also chiral odd and, moreover, T-odd, i.e., non-vanishing only due to final state 
interactions. Collins \cite{COL} was the first to propose such a spin dependent 
fragmentation function. It can be obtained, for example, in two-hadron production 
in $e^+e^-$ annihilation \cite{BJM}. In the cross section of SIDIS off transversely polarized 
nucleons it shows up as a $ \sin(\phi_h+\phi_S)$ dependence, where $\phi_h$ is the azimuthal 
angle of the outgoing hadron (with non-zero transverse momentum $P_{hT}$) around the 
virtual-photon direction, and $\phi_S$ is the azimuthal angle of the target spin vector, 
both in relation to the lepton scattering plane. 

\begin{figure}[b!] % fig 1
\epsfig{file=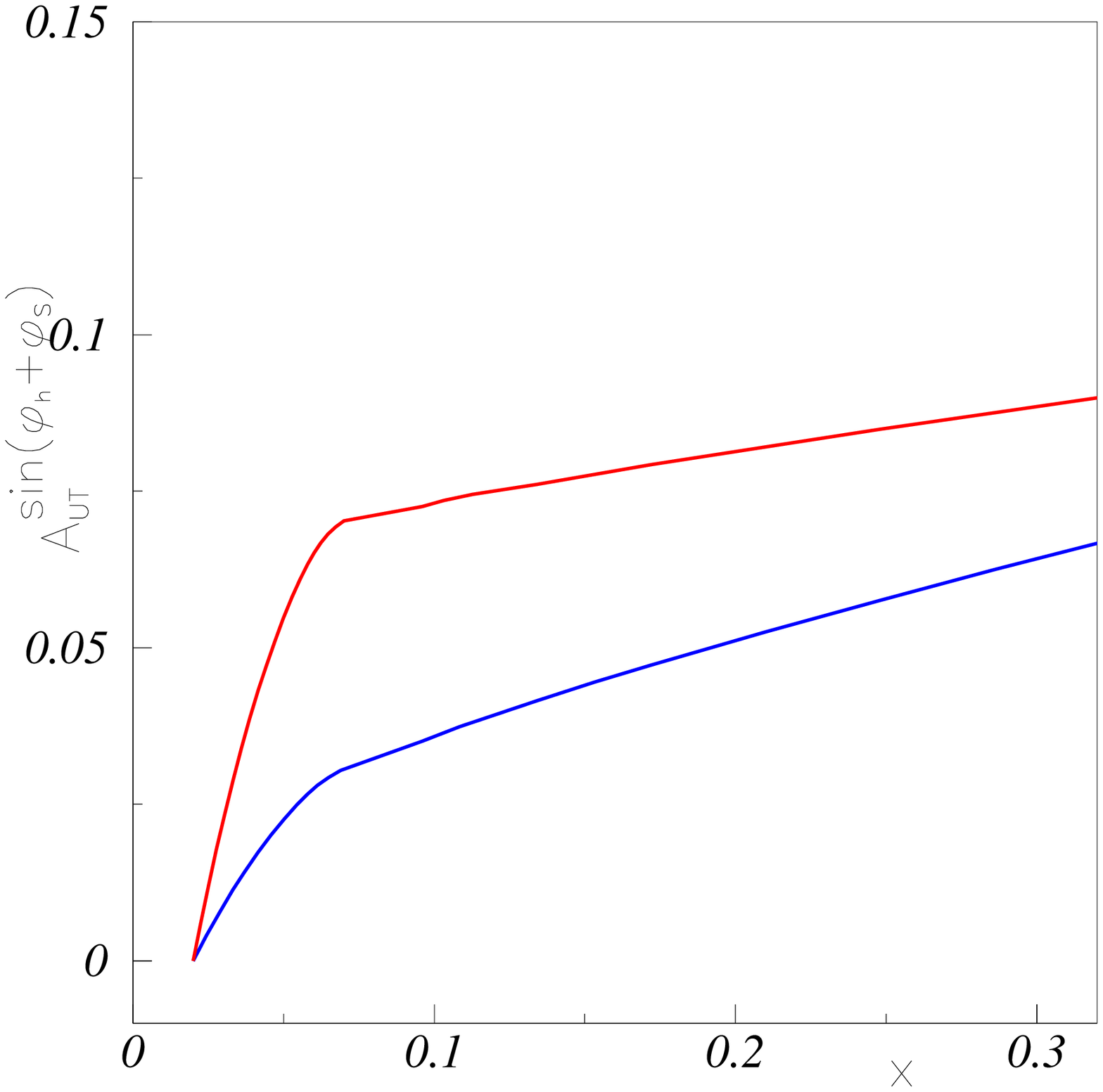,height=2.8in,width=2.8in}
\epsfig{file=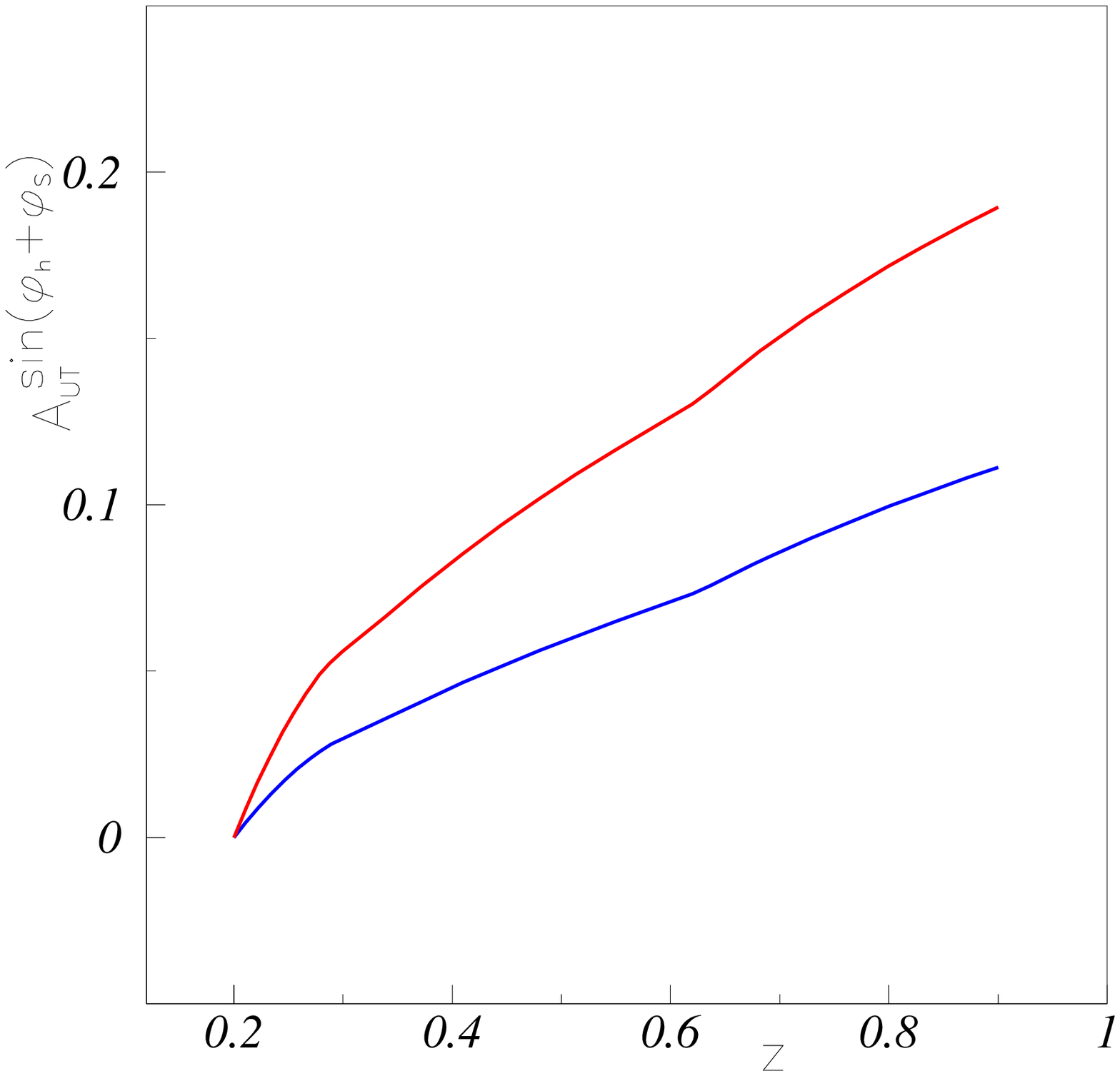,height=2.8in,width=2.8in}
\vspace{10pt}
\caption{The single target-spin asymmetry $A^{\sin(\phi_h+\phi_S)}_{UT}$ for 
$\pi^{+}$ production as a function of $x$ and $z$, evaluated using 
$M_C=2 m_{\pi}$ and $\eta=0.8$ in Eq.(\ref{AS3}). The two curves correspond   
to $h_1=g_1$ (lower curve) and $h_1=(f_1+g_1)/2$ (upper curve).}
\label{fig1}
\end{figure} 

The {\it sin}$(\phi_h+\phi_S)$ moment in the SIDIS 
cross-section can be related to the parton distribution and fragmentation 
functions involved in the parton level description of the underlying 
process \cite{AK,TM}. 
This moment is defined as the appropriately weighted integral over 
$P_{hT}$ (the transverse momentum of the observed hadron) of 
the cross section asymmetry:
\begin{equation}
\langle \frac{{\vert P_{hT}\vert}}{M_h} \sin (\phi_h+\phi_S) \rangle_{UT} \equiv 
\frac{\int d^2P_{hT} \frac{{\vert P_{hT}\vert}}{M_h}
\sin (\phi_h+\phi_S) \left(d\sigma^{\uparrow}-d\sigma^{\downarrow}\right)} 
{{1 \over 2} \int d^2P_{hT} \left(d\sigma^{\uparrow} + d\sigma^{\downarrow}\right)}, 
\label{AS}
\end{equation}
where $\uparrow (\downarrow)$ denotes the up (down) 
transverse polarization of the target in the virtual-photon frame,  
$M$ ($M_h$) is the mass of the target (final hadron), and 
the subscripts $U$ and $T$ indicate unpolarized beam and transversely  
polarized target, respectively.  

This asymmetry is given by~\cite{COL,AK,TM}: 
\begin{equation}
\langle \frac{\vert P_{hT}\vert}{M_h} \sin (\phi_h+\phi_S) \rangle_{UT} = 
4S_T \frac{(1-y)\,h_1(x) z H_1^{\perp (1)}(z)}{(1+(1+y)^2) f_1(x) D_1(z)}.  
\label{AS1}
\end{equation} 

This weighted single target-spin asymmetry is related to 
the unweighted one through the following relation:  
\begin{equation}
A^{\sin(\phi_h+\phi_S)}_{UT} 
\approx {M_h \over {\langle P_{hT}\rangle }} \langle \frac{ 
P_{hT}}{M_h} \sin (\phi_h+\phi_S) \rangle_{UT}.  
\label{AS3}
\end{equation}
Note that in the case of a longitudinally polarized target this asymmetry 
gives contribution to the {\it sin}$\phi_h$ asymmetry~\cite{OABK}, which was 
recently observed in semi-inclusive deep inelastic lepton scattering off a 
longitudinally polarized proton target at HERMES~\cite{HERM}. Describing this 
data by an approach where the transverse quark spin distribution in the 
longitudinally polarized nucleon is vanishing~\cite{SNO}, only about $25\%$ 
are contributed to {\it sin}$\phi_h$ by  the `kinematic` term that is 
proportional to the transverse component of the nucleon spin vector with respect to  
the virtual-photon momentum. 

For the numerical evaluation of the unweighted asymmetry $A_{UT}^{\sin (\phi_h+\phi_S)}$ 
the non-relativistic approximation 
$h_1(x) = g_1(x)$ is used as a lower limit and $h_1(x)=(f_1(x)+g_1(x))/2$ as 
an upper limit~\cite{SOF}. For the sake of simplicity, $Q^2$-independent 
parameterizations were chosen for the distribution functions $f_1(x)$ and 
$g_1(x)$~\cite{BBS}. 

To obtain the T-odd fragmentation function $H_1^{\perp (1)}(z)$,
the Collins ansatz~\cite{COL} for the analyzing power of 
transversely polarized quark fragmentation was adopted:
\begin{equation}
A_C(z,k_T) \equiv \frac{\vert k_T \vert}{M_h}\frac{H_1^{\perp}(z,k_T^2)}
{D_1(z,k_T^2)} = \eta \frac{M_C\,\vert k_T \vert}{M_C^2+k_T^2},
\label{H1T}
\end{equation}
where $\eta$ is taken as a constant, although, in principle it could be 
$z$-dependent, and $M_C$ is a typical hadronic mass whose value ranges from
$2m_{\pi}$ to $M_p$. 

In our calculations we use $M_C=2m_\pi$ and $\eta=0.8$ as in Ref.~\cite{OBSN}, 
where a good agreement was found with the single spin asymmetries of the 
distribution in the azimuthal angle $\phi_h$ for semi-inclusive $\pi^{+}$ production on a 
{\it longitudinally} polarized hydrogen target observed at HERMES~\cite{HERM,DELIA}. 
 
For the distribution of the final parton intrinsic transverse momentum, 
$k_T$, in the unpolarized fragmentation function $D_1(z,k^2_T)$, a Gaussian 
parameterization was used~\cite{KM} with $\langle z^2 k_T^2 \rangle = b^2$ 
(in the numerical calculations $b = 0.36$ GeV was taken~\cite{PYTHIA}). 
For $D_1^{\pi^{+}} (z)$, the parameterization from Ref.~\cite{REYA} was 
adopted.

In Fig.~\ref{fig1}, the asymmetry $A^{\sin(\phi_h+\phi_S)}_{UT}$ of 
Eq.(\ref{AS3}) for $\pi^{+}$ production on a transversely polarized 
proton target is presented as a function of $x$ and $z$. The curves 
have been calculated by integrating over the HERMES kinematic range 
taking $\langle P_{hT} \rangle = 0.365$ GeV as input. The latter 
value is obtained in this kinematic region assuming a Gaussian 
parameterization of the distribution and fragmentation functions 
with $\langle p_T^2 \rangle=(0.44)^2$ GeV$^2$~\cite{PYTHIA}. 

From Fig.~\ref{fig1} one sees  that the single transverse-target-spin 
asymmetry is quite large. In the HERMES kinematics $(\langle x 
\rangle \approx 0.1, \langle z \rangle \approx 0.4$) it amounts to $(4 \div 7)\%$. 
The HERMES experiment using a transversely polarized proton target will 
be able to extract $h_1(x)$ in a simple way, e.g. as suggested in Ref.~\cite{KNO}. 
First results on transverse quark spin distribution can be expected from 
combined results of HERMES and COMPASS within 3-5 years from now, while a complete 
high precision mapping of their $Q^2$- and $x$-dependence requires next-generation 
facilities, such us TESLA-N~\cite{TESLAN}, ELFE~\cite{ELFE}, eRHIC, EPIC, with high 
statistics measurements.

\end{document}